\documentclass{article}
\usepackage{spconf,amsmath,graphicx,amssymb,subfig}

\title{VITS-Based Singing Voice Conversion Leveraging Whisper and multi-scale F0 Modeling}
\name{Ziqian Ning$^{1}$, Yuepeng Jiang$^{1}$, Zhichao Wang$^{1}$, Bin Zhang$^{2}$, Lei Xie$^{1*}$\thanks{* Corresponding author.}}
\address{
$^1$Audio, Speech and Language Processing Group (ASLP@NPU), School of Computer Science, \\ Northwestern Polytechnical University, Xi'an, China\\
  $^2$Lyra Lab, Tencent Music Entertainment, Shenzhen, China
}
\begin{document}
%
\maketitle
\begin{abstract}
This paper introduces the T23 team's system submitted to the Singing Voice Conversion Challenge 2023. Following the recognition-synthesis framework, our singing conversion model is based on VITS, incorporating four key modules: a prior encoder, a posterior encoder, a decoder, and a parallel bank of transposed convolutions (PBTC) module. We particularly leverage Whisper, a powerful pre-trained ASR model, to extract bottleneck features (BNF) as the input of the prior encoder. Before BNF extraction, we perform pitch perturbation to the source signal to remove speaker timbre, which effectively avoids the leakage of the source speaker timbre to the target. Moreover, the PBTC module extracts multi-scale F0 as the auxiliary input to the prior encoder, thereby capturing better pitch variations of singing. We design a three-stage training strategy to better adapt the base model to the target speaker with limited target speaker data. Official challenge results show that our system has superior performance in naturalness, ranking 1st and 2nd respectively in Task 1 and 2. Further ablation justifies the effectiveness of our system design.


\end{abstract}
\begin{keywords}
Singing voice conversion, VITS, F0-modeling
\end{keywords}
\section{Introduction}
\label{sec:intro}

Voice conversion (VC) aims to convert the speech of a source speaker to that of another speaker while maintaining the linguistic content and speaking style. To this end, the main idea of VC is to disentangle speech into multiple factors, including speaker timbre, linguistic content, and speaking style, and then the linguistic content and speaking style are combined with the desired timbre of the target speaker to deliver the target speech. Similarly, singing voice conversion (SVC) mainly focuses on the conversion of a singing voice. In SVC, modeling expressive singing styles, such as temporal pitch variations, is one of the crucial problems. Besides, since different speakers have different pitch ranges and singing styles, the converted singing voice following the original singing style may hurt the similarity to the target singer. Thus the level of disentanglement needs to be properly considered, and it is challenging to achieve both high singing naturalness and high speaker similarity in SVC.


In the early attempt of SVC, many studies~\cite{DBLP:conf/interspeech/SainoZNLT06,DBLP:conf/ssw/OuraMYMNT10,DBLP:conf/icassp/NakamuraONT14} have used parallel training data between the source singer and the target singer to learn a mapping between the parallel samples. Since parallel singing data of different singers is difficult to collect, non-parallel SVC approaches leveraging non-parallel training data have obtained wide attention. The key idea of non-parallel SVC is to decompose the linguistic content and singing style from the source singing voice, and then generate the target singing voice with the target singer timbre. Variational autoencoder (VAE)~\cite{DBLP:conf/icassp/LuoHAH20} and generative adversarial network (GAN)~\cite{DBLP:conf/apsipa/LuZS020} based approaches were investigated for non-parallel SVC to disentangle linguistic content in an unsupervised manner. Moreover, information bottleneck ~\cite{DBLP:conf/ijcnn/TakahashiSM21} was introduced to reduce the correlation between content, style, and singer timbre. Instead of learning disentanglement during SVC model training, an intuitive way is to extract representation of linguistic content from pre-trained models in prior, such as neural bottleneck features (BNF)~\cite{DBLP:conf/icmcs/SunLWKM16,DBLP:conf/icassp/ZhaoLSWKTM22} and self-supervise learning (SSL) features~\cite{DBLP:journals/taslp/HsuBTLSM21, jayashankar2023self,DBLP:conf/interspeech/WangLT0WYM22}, extracted from an automatic speech recognition (ASR) model and an SSL model, respectively. Since the crucial role of singing style in the SVC task, many studies~\cite{jayashankar2023self,DBLP:conf/icassp/DengYLW020} have considered modeling pitch to represent singing style. For example, in~\cite{DBLP:conf/icassp/ZhouL22a,DBLP:conf/icmcs/LiuCH0M21,DBLP:conf/interspeech/LiLS22}, pitch information has been modeled in multi-temporal granularity with multiple network layers. Besides, style can be modeled specifically by a reference encoder to represent the singing style~\cite{DBLP:conf/interspeech/WangLT0WYM22}. 



As the fourth edition of the Voice Conversion Challenge (VCC), the recently launched Singing Voice Conversion Challenge (SVCC) has particularly focused on singing conversion~\cite{DBLP:journals/corr/abs-2306-14422}, providing a common testbed to benchmark different SVC approaches. Similar to the past VCC iterations, the primary goal is to conduct speaker/singer conversion. Specifically, SVCC2023 is separated into two any-to-one tasks: \textit{in-domain} SVC and \textit{cross-domain} SVC. For the in-domain SVC, the main task is to convert a source singing clip to that of a target speaker using the singing voices of the speaker as training data, while the cross-domain SVC poses a greater challenge as only speech data is available for the target speaker. 
It is worth noticing that only about 10 minutes of speaking/singing data is provided for each target speaker, positioning the task as a low-resource one. The data usage is limited to a list of open-source repositories according to the rules~\footnote{\label{svcc}http://vc-challenge.org/rules.html}.






\begin{figure}[ht]
\centering
\includegraphics[scale=0.65]{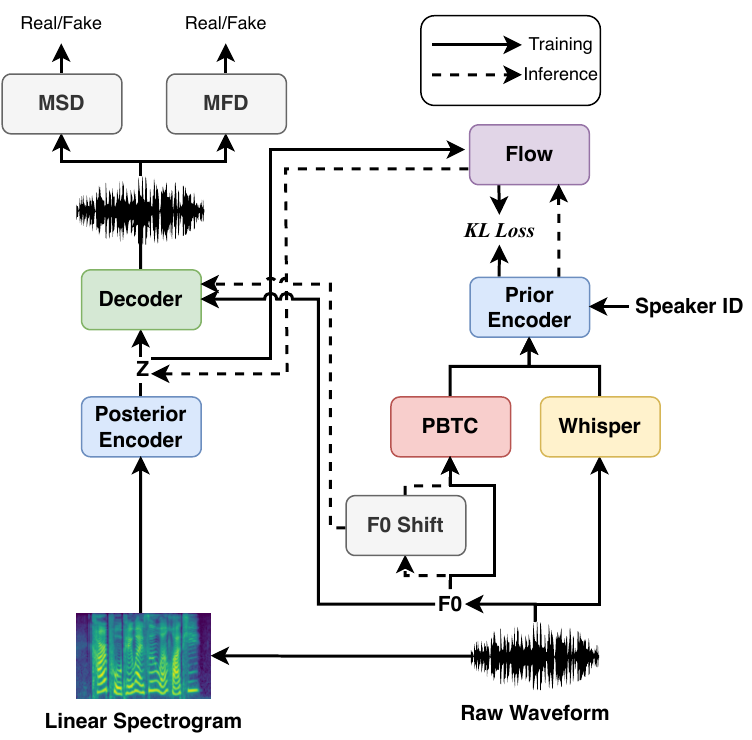}
\caption{The architecture of the proposed system}
\label{fig:overall}
\end{figure}


In this paper, we present our SVC system for both in- and cross-domain tasks of SVCC 2023. Our conversion model is based on VITS~\cite{DBLP:conf/icml/KimKS21}, consisting of a posterior encoder, a prior encoder, and a decoder. Following the recognition-synthesis framework, we first train a base model using the open-source data repositories and then fine-tune the model using the target speaker data. To obtain a robust content representation from input audio, we leverage a powerful ASR model -- Whisper~\cite{DBLP:journals/corr/abs-2212-04356} -- to perform BNF extraction. Inspired by a previous study~\cite{DBLP:journals/corr/abs-2210-15158}, we extract BNF from the shallow encoder layers of Whisper, in which the BNF is believed to contain not only linguistic content but also rich style-related information such as prosody. Moreover, speech perturbation is first adopted to the source speech before BNF extraction to remove speaker timbre and thus prevent its leakage to the target speech. Besides, to better capture pitch variations of the source singing voice, parallel bank of transposed convolutions (PBTC) module~\cite{jayashankar2023self,DBLP:conf/icassp/ZhouL22a} is introduced to extract multi-scale F0 from different temporal granularity. The F0 features are thus fed into the prior encoder along with the BNF.  We further introduce a three-stage training strategy, including warm-up, pre-training, and adaptation, to better optimize the model. Official challenge results show that our system has superior performance in naturalness,
ranking 1st and 2nd respectively in Task 1 and 2~\cite{DBLP:journals/corr/abs-2306-14422}. Further ablation justifies the effectiveness of our system design. Audio samples can be found on our demo page.\footnote{https://nzqian.github.io/SVCC2023-t23-ASLP/}


\section{proposed system}
\label{sec:format}

\subsection{System Overview}
As depicted in Fig~\ref{fig:overall}, the proposed system is based on VITS~\cite{DBLP:conf/icml/KimKS21}, consisting of four key components: a posterior encoder, a prior encoder, a PBTC module, and a decoder. The concatenation of the posterior encoder and decoder acts as a self-reconstruction manner to model the waveform as a hidden representation and then predict the origin waveform. The prior encoder fuses singer timbre, singing style, and linguistic content. Furthermore, an invertible flow is adopted to bridge the prior encoder and posterior encoder.  

\textbf{Posterior Encoder:} Following the architecture of VITS, we employ non-causal WaveNet~\cite{DBLP:conf/ssw/OordDZSVGKSK16} residual blocks in the posterior encoder. By utilizing the linear spectrogram extracted from the original singing waveform $y$ as input, the posterior encoder's objective is to model the posterior distribution $p(z|y)$ of hidden representation $z$.


\textbf{Prior Encoder:} The prior encoder is implemented using a multi-layer Transformer~\cite{DBLP:conf/nips/VaswaniSPUJGKP17}. Given the BNFs and multi-scale F0 extracted by Whisper and the PBTC module, denoted as $c_{bnf}$ and $c_{f0}$ respectively, the prior encoder, followed by a flow, estimates the prior distribution $p(z|c_{bnf}, c_{f0}, singer)$ with the target singer timbre. In order to bridge the distribution between the prior encoder and the posterior encoder, a normalizing flow is introduced to perform an invertible transformation of a simple distribution into a more complex distribution.


\textbf{Decoder:} The decoder is responsible for generating the singing waveform using the extracted latent representation $z$. Different from the HIFI-GAN decoder in VITS, to improve the singing voice reconstruction quality, we extend the decoder to the neural source filter (NSF) scheme~\cite{DBLP:journals/taslp/WangTY20}. Specifically, the extended decoder consists of a source module and a filter module. The source module transforms the F0 to a sine-based excitation signal, which can be defined as:
\begin{equation}
e_t=\left\{
\begin{aligned}
&\alpha sin(\sum_{k=1}^t 2\pi \frac{f_k}{N_s} + \phi) + n_t , & f_t > 0 \\
&100n_t , & f_t = 0
\end{aligned}
,\right.
\end{equation}
where $n_t \sim \mathcal{N}(0,0.003^2)$, $\phi \in [-\pi, \pi]$ is a random initial phase and $N_s$ is the waveform sampling rate. The filter module achieved by the HIFI-GAN decoder fuses the excitation signal and intermediate representation $z$ into the target singing waveform. In line with VITS, the discriminator, which comprises both a multi-period discriminator (MPD) and a multi-scale discriminator (MSD), is employed to enhance the quality of the reconstructed singing voice further.

The design of BNF extraction and F0 modeling are introduced in the following sections. Besides, to make a better adaptation performance, several training strategies are also illuminated in Section~\ref{sec:TS}.

\subsection{Bottleneck Feature Extraction}


Obtaining well-represent linguistic content is crucial in non-parallel SVC, as it directly impacts the intelligibility of converted results. Given that the datasets provided in SVCC 2023 comprise singing and speech recordings in different languages, relying solely on English datasets would be insufficient to train an SVC model. To address this challenge, a multi-lingual ASR is needed to perform BNF extraction. Recently, 
Whisper~\cite{DBLP:journals/corr/abs-2212-04356}, which is trained on 680,000 hours of multilingual speech data, has demonstrated high recognition accuracy and robustness across multiple languages. Hence, we believe that Whisper is a reasonable choice to obtain robust BNF on our multilingual training data.

Drawing inspiration from Chen et al.~\cite{DBLP:journals/corr/abs-2210-15158}, we observe that BNF from different ASR encoder layers shows variation effects on VC results in terms of speech intelligibility, style correlation, and speaker similarity. Thus, we use the BNF from the shallow encoder layer to ensure the high-fidelity linguistic content contained in BNF. And the rich style-related information contained in BNF from the shallow layer also contributes to the singing style modeling, which is crucial for the SVC task. Furthermore, to prevent the speaker-related information in BNF from leaking to converted results, we introduce random pitch perturbations~\cite{DBLP:conf/nips/ChoiLKLHL21} to the singing waveform before BNF extraction.

\begin{figure}[ht]
\centering
\includegraphics[scale=0.53]{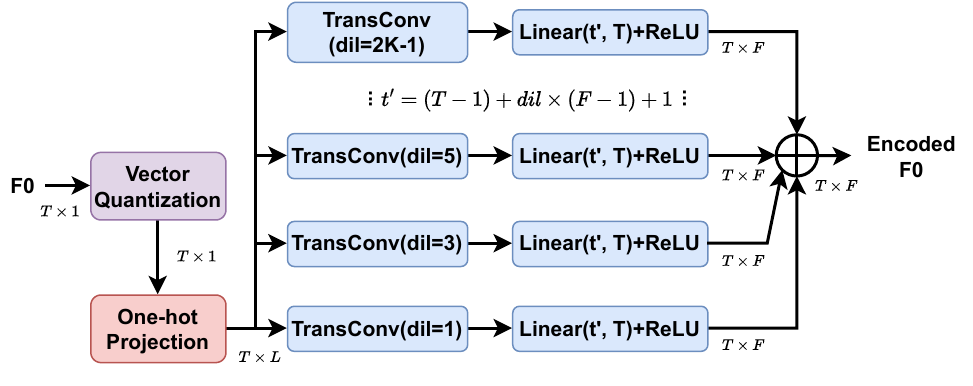}
\caption{The PBTC module for multi-scale F0 modeling}
\label{fig:PBTC}
\end{figure}

\subsection{Multi-scale F0 Modeling}
The singing style plays a crucial role in the SVC task, directly influencing the naturalness of the converted singing voice. Similar to many previous studies~\cite{DBLP:conf/interspeech/WangLT0WYM22}, we aim to represent the singing style of the source singing voice using F0. However, directly applying F0 sometimes causes unexpected results, such as unnatural singing and out-of-tone, primarily due to errors in the F0 estimation algorithm. Moreover, the SVC model is hard to capture the style relation among different timesteps without a specific-design structure. To address this challenge, we incorporate a parallel bank of transposed convolutions (PBTC) module~\cite{DBLP:conf/interspeech/WebberPK20,jayashankar2023self,DBLP:conf/icassp/ZhouL22a} to extract multi-scale F0 from multi-temporal granularity.

As shown in Fig.~\ref{fig:PBTC}, the PBTC module comprises a vector quantization module, a projection layer, and an array of 1D transposed convolutional layers. Each of these convolutions utilizes a different dilation rate, followed by a linear layer. Specifically, the F0 sequence is firstly globally normalized and quantized into $L$ bins. The resulting quantized F0 is then one-hot encoded and projected by the Linear projection layer. $K$ dilated transposed convolution layers with $F$ filters and different dilate rates then separately process the quantized F0, which enables to summarize F0 information from multi-temporal granularity. The resulting outputs with different time $t'$ are projected to time $T$ on the temporal axis to match the original duration before being fed to the prior encoder.

\subsection{Training Strategy}
\label{sec:TS}
\subsubsection{Three-stage Training Procedure}
In general, to effectively learn low-resource singers with only a few utterances, the SVC model is pre-trained on multi-speaker singing data and then adapts to the target singer. However, given the limited singing data available in SVCC and considering that the amount of data used for training a VITS-like model directly affects reconstruction quality and robustness, we propose an additional warm-up stage. In this stage, we utilize speech data to initially train the SVC model before transitioning to pre-training and adaptation.  The training procedure can be concluded as: 
\begin{itemize}
    \item[1)] Warm-up: init the SVC model on speech data.
    \item[2)] Pre-training: train the SVC model in singing data.
    \item[3)] Adaptation: adapt to the target singer.
\end{itemize}

\subsubsection{Data Augmentation}
Adapting the SVC model to the target speaker may encounter model overfitting as the target speaker data is quite limited. Therefore, in order to alleviate this problem, we augment the training data of the target singer to increase both the data quantity and diversity. The four data augmentation functions utilized are  formant shifting, pitch randomization, random frequency, and speed adjustment. These augmentation functions contribute to a larger and more diverse training data of the target singer, enhancing the model's ability to make the target speaker sing in different styles. It also alleviates the problem of speaker-related information leakage due to using shallow BNF mentioned above.




\subsubsection{Loss Function}

Following VITS~\cite{DBLP:conf/icml/KimKS21}, our model can be represented as a conditional VAE (CVAE) aiming to maximize the variational lower bound, also known as the evidence lower bound (ELBO). Consequently, the training loss of CVAE is the negative ELBO, comprising two main components: the reconstruction loss $\mathcal{L}_{recon}$ and KL loss $\mathcal{L}_{KL}$. The loss of CVAE can be described as:
\begin{equation}
    \mathcal{L}_{KL} = \mathcal{D}_{KL}(q(z|y)\Vert p(z|c)),
\end{equation}
\begin{equation}
    \mathcal{L}_{cvae} = \mathcal{L}_{recon} + \mathcal{L}_{KL},
\end{equation}
where $c$ is the condition of CVAE comprising of BNFs, F0, and speaker ID. $L_{recon}$ denotes the L1 distance of the Mel spectrogram between the ground truth $y$ and the generated waveform $\hat{y}$. And $\mathcal{D}_{KL}$ is the Kullback-Leibler divergence calculated between prior and posterior distributions. To further enhance the quality of the reconstructed waveform, we employ GAN-based training. Following HiFi-GAN~\cite{DBLP:conf/interspeech/SuJF20}, the discriminators D consisting of both multi-period discriminator (MPD) and multi-scale discriminator (MSD) are introduced in our model. The GAN losses for the decoder $G$ and the discriminator $D$ are defined as:
\begin{equation}
L_{adv}(G) = \mathbb{E}_{(z)}[(D(G(z)) - 1)^2],
\end{equation}
\begin{equation}
L_{adv}(D) = \mathbb{E}_{(y, z)}[(D(y) - 1)^2 + (D(G(z)))^2].
\end{equation}

With CVAE and adversarial training described above, the final loss used in the warm-up and pre-training can be described as:
\begin{equation}
L(G) = L_{adv}(G) + L_{cvae},
\end{equation}
\begin{equation}
L(D) = L_{adv}(D).
\end{equation}

For the adaptation process, to further prevent the SVC model from overfitting to the target singer data, we also incorporate weight regularization~\cite{DBLP:conf/icassp/WangXLDXZB22}, specifically utilizing a variant of L2 regularization. The regularization term, denoted as $L_{wReg}$, is defined as:
\begin{equation}
L_{wReg} = \Vert\theta - \hat{\theta}\Vert^2,
\end{equation}
where $\theta$ represents the parameters of the model before adaptation, and $\hat{\theta}$ refers to the current model parameters. The main purpose of $L_{wReg}$ is to prevent the adapted model's parameters from deviating significantly from those learned on the large dataset, thereby promoting more robust adaptation results. 




\begin{table}[]
\centering
 \caption{The detail of dataset we used for SVCC 2023}
 \label{tab:dataset}
 \setlength{\tabcolsep}{1.0mm}
\renewcommand\arraystretch{1.3}
\begin{tabular}{l|ccc}
\hline
        & Speakers           & Language           & Duration (hours)                \\ \hline
VCTK~\cite{veaux2017cstr}     & 109          & English           & 44.0                 \\
NUS48E~\cite{DBLP:conf/apsipa/DuanFLSW13}  & 12          & English           & 2.8              \\
Opencpop~\cite{DBLP:conf/interspeech/WangWZWLXZXB22} & 1          & Mandarin           & 5.2                 \\ 
M4singer~\cite{DBLP:conf/nips/ZhangLWDL0HHZCZ22} & 20         & Mandarin            & 29.8                   \\
OpenSinger~\cite{DBLP:conf/mm/HuangC0LCZ21} & 66          & Mandarin           & 51.9                   \\
SVCC2023\footnote[1]{} & 4 & English  & 0.6 \\
\hline
\end{tabular}
\end{table}

\section{Evaluation Results}
\label{sec:typestyle}

\subsection{Dataset}
The datasets we used for SVCC 2023 are detailed in Table~\ref{tab:dataset}. To warm up our SVC model, we employ VCTK~\cite{veaux2017cstr}, which comprises 44 hours of English speech recordings. For the pre-training stage, we utilize a mixture of NUS48E~\cite{DBLP:conf/apsipa/DuanFLSW13}, Opencpop~\cite{DBLP:conf/interspeech/WangWZWLXZXB22}, M4singer~\cite{DBLP:conf/nips/ZhangLWDL0HHZCZ22}, and OpenSinger~\cite{DBLP:conf/mm/HuangC0LCZ21}. This amalgamation consists of approximately 90 hours of singing data provided by 99 singers. For the in-domain SVC task, the training data consists of singing recordings from one male and one female singer, with each having 150 and 159 utterances, respectively. In contrast, the cross-domain SVC task involves a male and female target singer, but with 161 and 159 speech utterances, respectively. All audio samples in the dataset are uniformly sampled at a rate of 24 kHz, ensuring consistency across the experiments.

\subsection{Implementation Details}
Our system is built upon the open-source SVC project\footnote{https://github.com/svc-develop-team/so-vits-svc} but with substantial improvements described above. For BNF extraction, we extract 1024-dim BNF from the 20th layer of the medium-sized Whisper model\footnote{https://github.com/openai/whisper}. And we use PYIN~\cite{DBLP:conf/icassp/MauchD14} to extract F0 from the waveform. In our SVC system, F0 is quantized into $L=256$ bins, and the PBTC module consists of $K=10$ transposed convolution layers with $F=256$ filters. The posterior encoder, prior encoder, and decoder follow the same configuration in VITS~\cite{DBLP:conf/icml/KimKS21}. During training, we train the SVC model for 400k and 200K steps in the warm-up stage and pre-training stage, respectively, with a batch size of 96. And 50k training steps are performed in the adaptation stage. The learning rate is set to 1e-4, and the Adam optimizer ($\beta_1$ = 0.8, $\beta_2$ = 0.99) is used to optimize the SVC model. The speech augmentation used in adaptation following the coefficients setting of NANSY~\cite{DBLP:conf/nips/ChoiLKLHL21}. For inference, we implement a simple F0 shifting strategy to transform the distribution of F0 according to the mean and variance of F0 from the source singing voice and the target singer.


\subsection{Subjective Evaluation of SVCC 2023}

\subsubsection{Evaluation Metrics}
The typical 5-point scale mean opinion score (MOS) is used to measure naturalness subjectively. And for similarity, the same/different paradigm is employed with a 4-point scale, in which listeners are asked to choose from four options: (1) different singer, absolutely sure, (2) different singer, not sure, (3) same singer, not sure, (4) same singer, absolutely sure, between converted utterance and target singer utterance. 

To assess the SVC performance, crowd-sourced perceptual evaluations are conducted separately for English and Japanese listeners, yielding a total of 12,720 scores from English listeners and 38,160 scores from Japanese listeners. Each system includes an average of 120 scores from English listeners and 360 scores from Japanese listeners.

\subsubsection{Evaluation Results}
\begin{figure}[ht]
\centering
\includegraphics[scale=0.29]{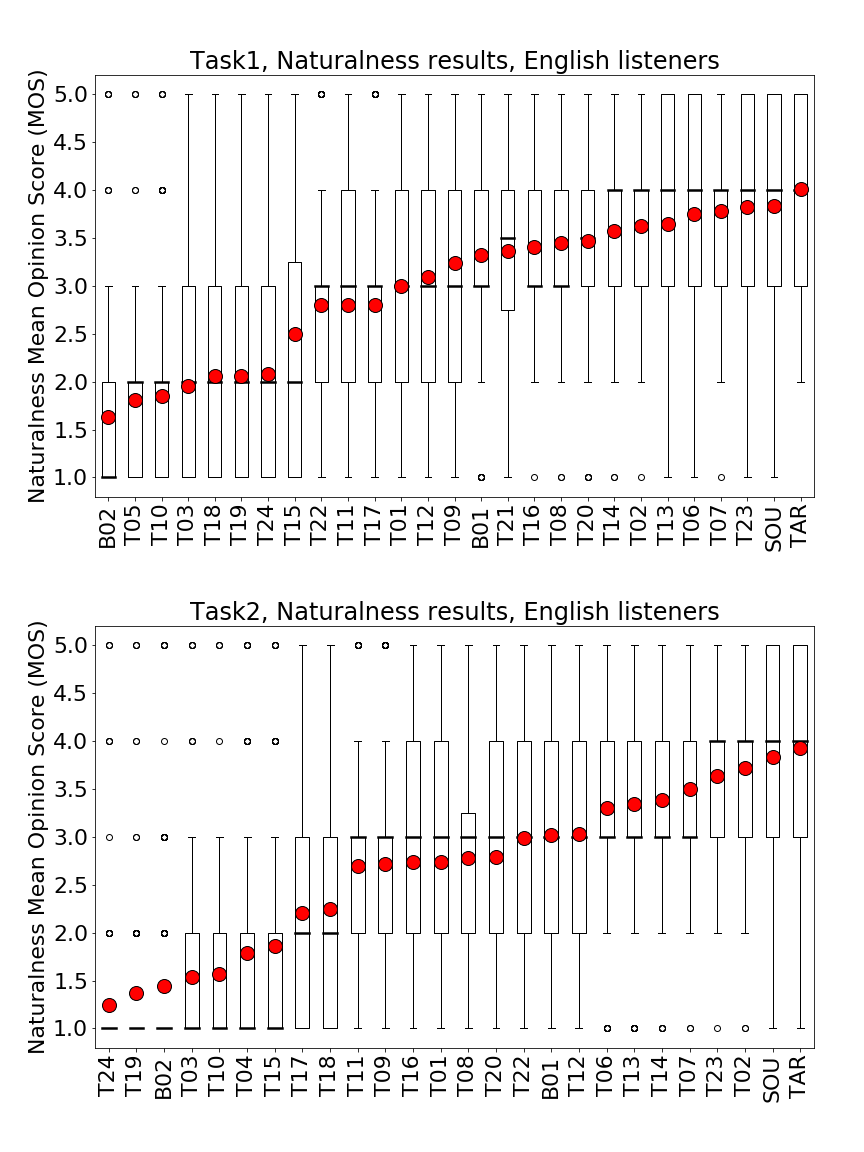}
\caption{Average naturalness MOS results of different teams for both in-domain (Task1) and cross-domain VC (Task2). Our team is T23.}
\label{fig:nat}
\end{figure}

\begin{figure}[ht]
\centering
\includegraphics[scale=0.30]{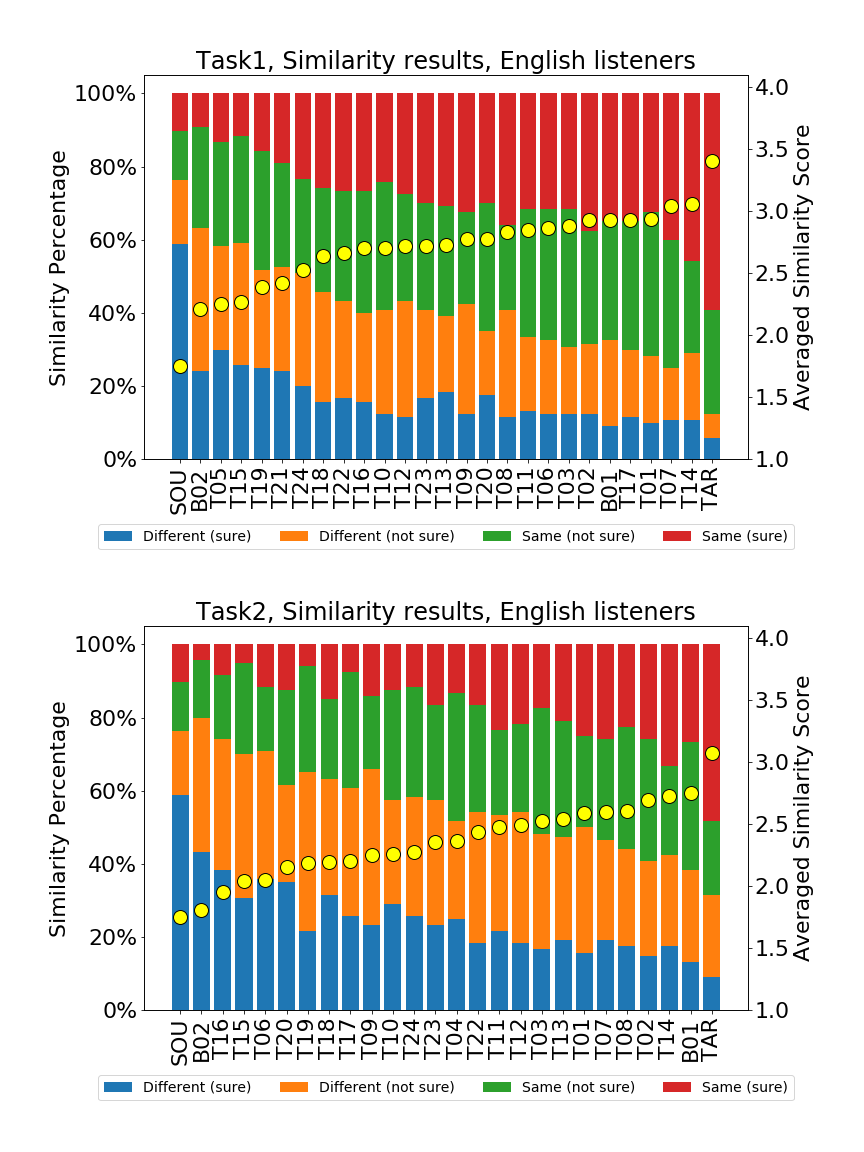}
\caption{Average similarity MOS results of different teams for both in-domain (Task1) and cross-domain VC (Task2). Our team is T23.}
\label{fig:sim}
\end{figure}

In SVCC 2023, a total of 24 teams participated. Among the participating systems, baseline systems B01 (DiffSVC~\cite{DBLP:conf/asru/LiuCSM21}) and B02 (FastSVC~\cite{DBLP:conf/icmcs/LiuCH0M21}) were provided by the organizers. Our system denoted as T23, was submitted for both tasks. Regarding naturalness and similarity, the evaluation results conducted by English listeners are presented in Fig.~\ref{fig:nat} and Fig.~\ref{fig:sim}, respectively.

\textbf{Naturalness:} As depicted in Fig.~\ref{fig:nat}, our system (T23) ranks 1 and 2 in Task 1 and 2, respectively. The results indicate that our converted samples exhibit minimal differences from the ground truth samples, showcasing our system's ability to achieve human-level naturalness.

\textbf{Similarity:} 
Based on the similarity results in Fig.~\ref{fig:sim}, our system obtains the middle place among all submitted systems.
However, it is essential to note that the difference between our system and the top-ranked system in terms of similarity percentage is relatively small. We observe that while many systems achieve a level of naturalness close to real recordings, they fell short in similarity to the target singer. This observation aligns with that in the offical challenge summary~\cite{DBLP:journals/corr/abs-2306-14422}. Consequently, addressing the challenge of enhancing similarity in the SVC task warrants further investigation.


\begin{table}[]
\centering
 \caption{Ablation Study.}
\setlength{\tabcolsep}{0.5mm}
 \label{tab:ablation}
\setlength{\tabcolsep}{2mm}{
\begin{tabular}{l|ccc}
\hline
        & Naturalness                      & Similarity                \\ \hline
Proposed Model    & 3.85 $\pm$0.02                     & 3.47 $\pm$0.04                \\
\hspace{1em}w/o PBTC  & 3.62 $\pm$0.02                    & 3.43 $\pm$0.05             \\
\hspace{1em}Typical Conformer ASR & 3.58 $\pm$0.03                    & 3.53 $\pm$0.02                \\ 
\hspace{1em}w/o Warm-up     & 3.55 $\pm$0.04                    & 3.40 $\pm$0.04                  \\ \hline
\end{tabular}
}
\end{table}

\subsection{Ablation Study} 
To evaluate the effectiveness of our system design, we conducted three ablations. First, we dropped PBTC to show the significance of multi-scale F0 modeling, denoted as \textit{w/o PBTC}. Second, we replaced F0 with a typical multilingual conformer-based ASR trained on WenetSpeech ~\cite{DBLP:conf/icassp/ZhangLGSYXXBCZW22} and LibriSpeech ~\cite{DBLP:conf/icassp/PanayotovCPK15}, referred to as \textit{Typical Conformer ASR}. Third, we removed the warm-up stage from our three-stage training strategy, denoted as \textit{w/o Warm-up}.
The ablation study comprised 30 singing clips from all four target singers, and it was evaluated by 20 listeners.
As shown in Table~\ref{tab:ablation}, the exclusion of the PBTC module resulted in noticeable degradation of naturalness. Similarly,  replacing the BNF feature also led to a decline in naturalness, although it resulted in slightly higher similarity scores. Without the warm-up stage, the SVC model lacked sufficient training for robust generation, leading to a reduction in naturalness.



\section{Discussion \& Conclusions}
\label{sec:majhead}
In evaluating the competition results, we observed that the converted singing clips demonstrated remarkable naturalness, closely resembling real recordings. However, assessing similarity presented challenges due to the diverse singing styles across different individuals and songs. Additionally, in the cross-domain task, accurately determining the speech speaker's timbre during singing proved to be even more challenging. Notably, we observed differences in energy levels between our conversion results and the ground truth, which prompts further investigation into its impact on the listener's perception.

In this paper, we present our SVC system for the Singing Voice Conversion Challenge 2023. Building upon the architecture of VITS, our SVC system consists of a posterior encoder, a prior encoder, a decoder, and a PBTC module. Following the recognition-synthesis framework, the SVC model uses BNF extracted from the shallow encoder layer of Whisper as content representation and fuses it with the source singing style and target speaker timbre to generate the target singing voice. In addition to using F0 to represent singing style, the PBTC module can extract multi-scale pitch information from F0 to better capture the pitch variation of the source singing waveform. Moreover, we designed a three-stage training strategy to ensure model optimization and enhance the model's adaptation ability. Official challenge results of our model in
SVCC2023 demonstrate superior performance in naturalness.



\bibliographystyle{IEEEbib}
\bibliography{strings,refs}

\end{document}